\documentclass[prl,aps,twocolumn,twoside,superscriptaddress]{revtex4}

\usepackage{amssymb}
\usepackage{amsmath}
\usepackage{amscd}
\usepackage{latexsym}
\usepackage{epsfig}
\usepackage{graphicx}
\usepackage{bbm}

\newtheorem{defi}{Definition}

\newtheorem{exempel}[defi]{Example}

\def\CA{{\cal A}}

       \def\CN{{\cal N}}       
              
\def\CS{{\cal S}}              \def\CU{{\cal U}}

\def\beq{\begin{equation}}
\def\eeq{\end{equation}}

\def\<{\langle}
\def\>{\rangle}

\newcommand{\tr}{{\operatorname{Tr}}}
\newcommand{\id}{{\operatorname{id}}}

\newcommand{\ket}[1]{{|{#1}\rangle}}

\newcommand{\1}{{\openone}}

\def\proj#1{ | #1 \rangle \langle #1 |}

\newlength{\blank}
\newlength{\equalsign}
\begin{document}

\title{A triangle of dualities: reversibly decomposable quantum channels, source--channel
duality, and time reversal}
\date{\today}
\author{I. Devetak}
\email{devetak@usc.edu}
\affiliation{
Electrical Engineering Department,
University of Southern California,
Los Angeles, CA 90089, USA}

\begin{abstract} 
Two quantum information processing protocols are said to be dual under 
resource reversal if the resources consumed (generated) in one protocol are 
generated (consumed) in the other. Previously known examples include 
the duality between entanglement concentration and dilution, and
the duality between coherent versions of teleportation and super-dense coding. 
A quantum feedback channel is an isometry from a system belonging
to Alice to a system shared between Alice and Bob.
We show that such a resource may be reversibly decomposed into a perfect 
quantum channel and pure entanglement, generalizing both of the
above examples. The dual protocols responsible for 
this decomposition are the ``feedback father'' (FF) protocol and 
the ``fully quantum reverse Shannon'' (FQRS) protocol.
Moreover, the ``fully quantum Slepian-Wolf'' protocol (FQSW),
a generalization of the recently discovered ``quantum state merging'', 
is related to FF by source-channel duality, and to FQRS by time reversal duality,
thus forming a triangle of dualities. The source-channel duality
is identified as the origin of the previously poorly understood 
``mother-father'' duality. Due to a symmetry breaking, the dualities
extend only partially to classical information theory.

\end{abstract}

\maketitle


The canonical example of an entangled state is an 
ebit, or EPR pair, 
$\Phi^{AB} = {1}/{\sqrt{2}} \, (\ket{0}^A \ket{0}^B + \ket{1}^A \ket{1}^B)$
shared between two spatially separated parties Alice and Bob. 
The systematic study of entanglement was initiated by the 
realization that a general pure 
bipartite state  $\ket{\phi}^{AB}$
is \emph{asymptotically equivalent} to a real number $E$ of ebits,
where $E = H(A)_\phi$, and $H(A)_\phi = -\tr \, \phi^{A}\log \phi^{A}$ 
is the von Neumann entropy of the restriction 
$\phi^{A} = \tr_B(\phi^{AB})$ of the state $\phi^{AB} = \proj{\phi}^{AB}$
to Alice's system. 
For any $\epsilon, \delta > 0$ and sufficiently large number $n$,
 there exists a protocol 
which transforms  $n$  copies of 
$\ket{\phi}^{AB}$ to a state that is $\epsilon$-close (say, in trace 
distance) to $\lfloor n(E - \delta)\rfloor$ ebits. This protocol
is called entanglement concentration \cite{BBPS}, 
and we can symbolically write the statement of its existence as  
a \emph{resource inequality} \cite{DHW}
$$
\< \phi \> \geq H(A)_\phi \, [q \, q].
$$
Here $\< \phi \>$ is the infinite sequence $(\phi^{\otimes n})_{n = 1}^{\infty}$.
The notation used for ebits $[q \, q]:= \< \Phi \>$ 
was introduced in \cite{DW}, along with corresponding notation for
qubit channels $[q \rightarrow q]$, classical bit channels $[c \rightarrow c]$
and bits of common randomness $[c \, c]$. $R \< \xi \>$ is defined
as $(\xi^{\otimes \lfloor R n \rfloor })_{n = 1}^{\infty}$.
In general we write an inequality $\geq$ between $(\phi_n)_{n = 1}^{\infty}$ and 
$(\psi_n)_{n = 1}^{\infty}$ if for any $\epsilon, \delta > 0$ and sufficiently 
large $n$  there is a protocol transforming 
$\phi_n$ into an $\epsilon$-approximation of $\psi_{\lfloor (1- \delta)n \rfloor}$.
As it turns out, the reverse is also true. The entanglement dilution  \cite{LP}
resource inequality reads 
$$
H(A)_\phi \, [q \, q] \geq \< \phi \>.  
$$
Dilution additionally consumes a sublinear amount  classical 
communication, but this corresponds to an asymptotic rate of $0$, and as
such does not enter into the resource count.
The two may be combined to give a resource \emph{equality}
\beq
\< \phi \> = H(A)_\phi \, [q \, q]. 
\label{ebitz}
\eeq
A single number $E$ thus suffices to characterize the asymptotic
properties of the state $\ket{\phi}^{AB}$.
Unfortunately, for more than two parties this appears to no longer hold. 
For instance, two GHZ states
$$
\Gamma^{RAB} = 1/\sqrt{2} \,
 (\ket{0}^{R} \ket{0}^{A} \ket{0}^{B} + \ket{1}^{R} \ket{1}^{A} \ket{1}^{B})
$$ 
cannot be reversibly decomposed into three pairs 
of ebits $\Phi^{RA}, \Phi^{RB}, \Phi^{AB}$, even though the local entropies of
$R$, $A$ and $B$ are all $2$ in both cases \cite{LPSW}.
\begin{figure}
  \includegraphics[width=8.5cm]{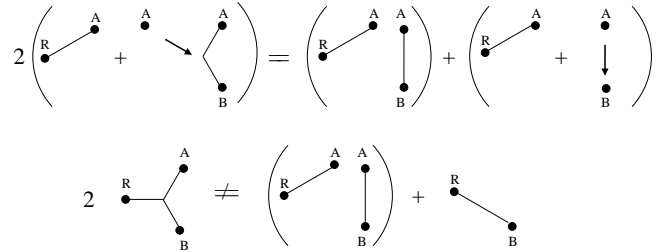}
\vspace{2.5 mm}  
\caption{A successful dynamic analogue of the failed GHZ decomposition. 
The first line is a graphical representation of (\ref{cobit})
with two EPR pairs $\Phi^{RA}$ added to both sides of the equality.
The second line represents the failure of the decomposition of two GHZ
states into three EPR pairs $\Phi^{RA}, \Phi^{RB}, \Phi^{AB}$.
The first line is a dynamic version of the second in the sense that
the LHS (RHS) of the latter is obtained by combining the resources on the
LHS (RHS) of the former.}
  \label{fig5}
\end{figure} 
However, a dynamic version of this decomposition does hold.
The construction of coherent versions of teleportation and super-dense coding leads
to the ``coherent communication'' resource equality \cite{aram}
\beq
2 [q \rightarrow qq] =  [q \rightarrow q] +  [q \, q].
\label{cobit}
\eeq
Here $[q \rightarrow qq]$ represents the 
coherent classical bit channel (or \emph{cobit}), an
isometry $\Delta: A' \rightarrow AB$ defined by
$$
\Delta: \ket{i}^{A'} \rightarrow \ket{i}^{A} \ket{i}^{B},  i = 0,1,
$$
where $\{ \ket{0}, \ket{1}\}$ is a preferred orthonormal basis of 
a qubit system. The relation to the GHZ decomposition problem is elaborated in figure 
\ref{fig5}.

The first result of this paper is a resource equality that generalizes  both 
(\ref{ebitz}) and (\ref{cobit}). 
A channel that creates general three-party pure states, in the same sense 
that cobits create GHZ states, can be reversibly decomposed into ebits 
$[q \, q]$ and qubits $[q \rightarrow q]$, and may thus be characterized by 
two parameters. The equality is a combination of two resource inequalities,
the ``fully quantum reverse Shannon'' (FQRS) and ``feedback father'' (FF) inequalities, 
which we describe below.

\paragraph{Fully quantum reverse Shannon inequality.} At this point
we need to introduce the concept of a 
\emph{relative resource}. Usually, when Alice and Bob are connected by 
a quantum channel $\CN: A' \rightarrow B$, no restriction is placed on Alice's 
input density operator, as long as it lives on a Hilbert space of the right dimension. 
For a fixed blocklength $n$, possessing a relative resource $\< \CN : \rho^{A'} \>$ 
means that only if Alice
inputs a density operator close to  $(\rho^{A'})^{\otimes n}$ 
is she guaranteed that the channel will behave like 
$\CN^{\otimes n}$. Relative resources come about naturally in the context of quantum 
compression. Using Schumacher compression \cite{Sch}, 
Alice is able to convey a good approximation to $n$ copies of some state 
$\rho^{A'}$ to Bob using $\approx n H(A')_\rho$ qubits of communication (for sufficiently
large $n$,  as usual). 
Letting $\varphi^{RA'}$ be a purification of $\rho^{A'}$ (i.e. a pure
state such that $\tr_R \varphi^{RA'} = \rho^{A'}$), 
this may be written as a resource
inequality
\beq 
H(R)_\varphi \, [q \rightarrow q] \geq
\< \id^{A' \rightarrow B} : \rho^{A'} \>,
\label{schu}
\eeq
i.e. we are able to simulate the identity channel $\id^{A' \rightarrow B}$, 
\emph{assuming} that the input density operator is close to
$(\rho^{A'})^{\otimes n}$.
The same simulation 
could never work for an input density matrix of a higher entropy, by the converse
to Schumacher's theorem \cite{Sch}.
What if one wishes to simulate an arbitrary channel 
$\CN: A' \rightarrow B$? The quantum reverse Shannon theorem 
\cite{QRST} gives
us a way to do this:
\beq
H(B)_\sigma \, [q \, q] + I(R; B)_\sigma \, [c \rightarrow c] \geq
\< \CN^{A' \rightarrow B} : \rho^{A'}  \>,
\label{qrsth}
\eeq
where 
$$
\sigma^{RB} = (\1^R \otimes \CN^{A' \rightarrow B}) \varphi^{RA'},
$$
and the \emph{quantum mutual information} is defined as
$I(R;B) = H(R) + H(B) - H(RB)$.
In fact, the protocol that achieves (\ref{qrsth}) accomplishes slightly more \cite{QRST}.
A noisy channel $\CN$ normally arises from  an isometry 
$\tilde{\CU}: A' \rightarrow BE$ with a larger target Hilbert space that includes
the unobserved environment $E$, followed by the tracing out of $E$.
In the simulation of $\CN$ the environment is also simulated, 
and ends up in Alice's possession at the end of the protocol.
Thus the channel we end up simulating is the \emph{quantum feedback channel} 
$\CU: A' \rightarrow AB$, an isometry from a system belonging
to Alice to a system shared between Alice and Bob,
and the resource inequality becomes
$$
H(B)_\sigma \, [q \, q] +  I(R; B)_\sigma \, [c \rightarrow c] \geq
\< \CU^{A' \rightarrow AB} : \rho^{A'}  \>.
$$
It is shown in \cite{DHLS} that the protocol can be made ``coherent''
\cite{aram, DHW}, yielding the \emph{fully quantum reverse Shannon} (FQRS) inequality
\beq
1/2  \, I(R;B)_\psi \, [q \rightarrow q] +  1/2  \, 
I(A;B)_\psi \, [q \, q] \geq \< \CU^{A' \rightarrow AB} : \rho^{A'}  \>,
\label{RI1}
\eeq
where 
\beq
\psi^{RAB} = (\1^R \otimes \CU^{A' \rightarrow AB}) \varphi^{RA'}
\label{dogz}
\eeq
is a purification of $\sigma^{RB}$.
Schumacher compression is  a special case of this inequality in 
which the feedback system $A$ is absent.

\paragraph{Feedback father.}
The ``father'' inequality  \cite{DHW} regards entanglement-assisted quantum communication
over the noisy quantum channel $\CN$:
\beq
\< \CN^{A' \rightarrow B} \>  +   1/2  \, I(R;A)_\psi \, [q \, q] \geq 
 1/2  \, I(R;B)_\psi \,  [q \rightarrow q].
\label{tata}
\eeq
The state $\psi^{RAB}$ is defined by (\ref{dogz}), noting that
entropic coefficients are independent of the choice of $\CU: A' \rightarrow AB$ 
for which $\CN = \tr_A \circ \CU$.
The first observation is that there is a protocol implementing 
(\ref{tata}) that merely requires the relative resource  
$\< \CN^{A' \rightarrow B} : \rho^{A'}\>$ instead of the full  
$\< \CN^{A' \rightarrow B} \> $,
$$
\< \CN^{A' \rightarrow B} : \rho^{A'}\>  +   1/2  \, I(R;A)_\psi \, [q \, q] \geq 
 1/2  \, I(R;B)_\psi \, [q \rightarrow q].
$$
The second observation is that if the feedback channel $\CU$ is given instead of the weaker
$\CN$, then applying the protocol from \cite{DHW} achieving (\ref{tata}),
Bob is left with a purification of Alice's system $A$, yielding an additional
$H(A) \, [q \, q]$ after entanglement concentration.
Thus 
\begin{eqnarray*}
\lefteqn{ \< \CU^{A' \rightarrow AB}  : \rho^{A'}\>  +   1/2 \,  I(R;A)_\psi \, [q \, q]}  \\
& \geq &  1/2  \, I(R;B)_\psi \, [q \rightarrow q] + H(A)_\psi \, [q \, q].
\end{eqnarray*}
Canceling terms and using 
\beq
H(A)_\psi = 1/2 \, I(R;A)_\psi + 1/2 \, I(A;B)_\psi,
\label{zub}
\eeq
gives the \emph{feedback father} (FF) inequality:
\beq
\< \CU^{A' \rightarrow AB}: \rho^{A'}\>  \geq 
 1/2  \, I(R;B)_\psi \, [q \rightarrow q] +  1/2  \, I(A;B)_\psi \, [q \, q].
\label{RI2}
\eeq
A special case of (\ref{RI2}) where there is no actual feedback is the
reverse of Schumacher compression:
\beq
\< \id^{A' \rightarrow B} : \rho^{A'} \> \geq 
H(R)_\varphi \, [q \rightarrow q].
\label{revschu}
\eeq

\begin{figure}
  \includegraphics[width=4cm]{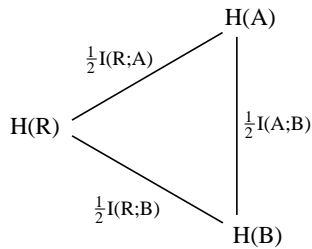}
  \caption{A hypothetical decomposition of (many copies of) the state $\psi^{RAB}$
into three entangled pure states. The values of the nodes are the local entropy rates.
The values of the links are the corresponding amounts of entanglement, 
which are uniquely determined by the local entropy values via (\ref{zub}). 
As in the GHZ case (see figure \ref{fig5}), it is only the dynamic version of this 
decomposition that holds.} 
  \label{fig3}
\end{figure}

\paragraph{Duality \#1: FQRS is related to FF  by resource
reversal.} Clearly, (\ref{RI1}) and  (\ref{RI2})  are reverses of each other, 
and together they give the resource equality
\beq
\< \CU^{A' \rightarrow AB}  : \rho^{A'}\>  =
 1/2  \, I(R;B)_\psi [q \rightarrow q] +  1/2  \, I(A;B)_\psi [q \, q].
\label{ricchi}
\eeq
A special case is (\ref{ebitz}) in which $\CU$ is the \emph{appending} channel
$\CA_\phi: A' \rightarrow A_1 A_2 B$, which relabels $A'$ as $A_1$,
and appends the state $\phi^{A_2 B}$ (i.e., it maps some
$\rho^{A'}$ to $\rho^{A_1} \otimes \phi^{A_2 B}$).
Another special case is (\ref{schu}) and (\ref{revschu}), in which $\CU$ is 
$\id^{A' \rightarrow B}$ and $A$ is null.
The third special case is (\ref{cobit}), where $\CU$ is the coherent classical bit 
channel $\Delta$, and $\rho$ is a maximally mixed qubit state $\tau$ 
(it can be shown that $\< \Delta: \tau \>$ is equivalent to $\< \Delta \>$).
Just as in the GHZ case, the naive static version of (\ref{ricchi}), which 
is not true, would be an asymptotic decomposition of $\psi^{RAB}$ 
into three maximally entangled states depicted in figure  \ref{fig3}.
It is satisfying to see these entropic coefficients reappear in the correct equation 
(\ref{ricchi}).

A resource equality of this generality can tell us a lot about optimal transformations 
between the resources involved, such as the capacity
of the quantum feedback channel $\CU^{A' \rightarrow AB}$ for simultaneous 
generation of quantum communication and entanglement. The task is to
find the set $\CS(\CU)$ of rate pairs $(Q,E)$ for which
$$
\< \CU^{A' \rightarrow AB}\> \geq
Q [q \rightarrow q] +  E [q \, q].
$$
The answer is given by
$\CS(\CU) = \bigcup_{n \rightarrow \infty} 1/n \, \CS^{(1)}(\CU^{\otimes n})$, where
$$
\CS^{(1)} = \bigcup_\psi \{ 1/2 (I(R;B)_\psi, 1/2 I(A;B)_\psi) \}. 
$$

\begin{figure}
  \includegraphics[width=8cm]{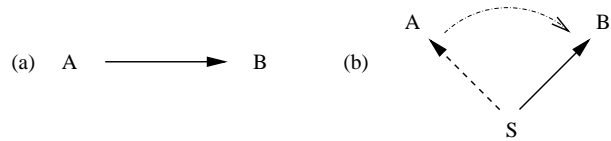}
  \caption{A channel (a) between Alice and Bob may be used in a source coding
problem (b) to convert the channel from the Source to Alice, into a channel
from the Source to Bob.} 
  \label{fig1}
\end{figure}

\paragraph{Fully quantum Slepian-Wolf} So far we have been dealing with what is traditionally
known as \emph{channel coding}: there are two parties Alice and Bob, and their
task is to effect conversions between resources, whether static, dynamic or relative.
In \emph{source coding} there is an additional protagonist, the Source. 
The Source holds a state purified by some reference system $R$.
Alice's and Bob's job is to effect conversions between 
{relative resources originating at the Source}.
A simple example of a source type resource inequality is
$$
\< \id^{S \rightarrow \hat{A}}: \rho^{S} \> +
\< \id^{A' \rightarrow B}: \rho^{A'} \> \geq 
\< \id^{S \rightarrow \hat{B}}: \rho^{S} \>,
$$
illustrated in figure \ref{fig1}. Combining it with Schumacher 
compression (\ref{schu}) gives
$$
\< \id^{S \rightarrow \hat{A}}: \rho^{S} \> 
+ H(S)_\rho \, [q \rightarrow q]
 \geq 
\< \id^{S \rightarrow \hat{B}}: \rho^{S} \>.
$$
While the two formulations are equivalent, 
compression is traditionally thought of in terms of source coding. 
More generally, the source may be initially distributed between Alice and Bob 
via a general isometry $\CU: S \rightarrow AB$, and the goal is to 
divert it entirely to Bob. A problem of this kind,
the quantum Slepian-Wolf problem, was first solved in \cite{HOW}.
The result of \cite{HOW} is generalized in \cite{ADHW} to
the \emph{fully quantum 
Slepian-Wolf} (FQSW) inequality, which reads
\begin{eqnarray}
\lefteqn{ \< \CU^{S \rightarrow AB} : \rho^S \> +  
1/2  \, I(R;A)_\psi [q \rightarrow q]} \nonumber \\ 
& \geq & 1/2  \, I(A;B)_\psi [q \, q] 
+ \< \id^{S \rightarrow \hat{B}} : \rho^{S} \>,
\label{RI3}
\end{eqnarray}
where 
$$
\psi^{RAB} = (\1^R \otimes \CU^{S \rightarrow AB}) \varphi^{RS}
$$
and $\varphi^{RS} = \proj{\varphi}^{RS}$ is a purification of $\rho^{S}$,
cf. (\ref{dogz}).

Since neither Alice nor Bob have control over the Source,
(\ref{RI3}) holds when applied to $\rho^{S}$, giving
the ``mother'' 
inequality \cite{DHW}
$$
 \< \sigma^{AB} \> +  
1/2  \, I(R;A)_\psi [q \rightarrow q] 
\geq 1/2  \, I(A;B)_\psi [q \, q], 
$$
which concerns quantum-communication-assisted distillation of entanglement from
$\sigma^{AB} =  \CU^{S \rightarrow AB}(\rho^{S})$.

\paragraph{Duality \#2: FF is related to FQSW via source-channel duality.}
The FF inequality (\ref{RI2}) may be combined with Schumacher compression 
(\ref{schu}), to give, after cancellation of terms,
\begin{eqnarray*}
\lefteqn{\< \CU^{A' \rightarrow AB} : \rho^{A'} \> +  1/2  \, 
I(R;A)_\psi [q \rightarrow q]} \\ 
& \geq &
1/2  \, I(A;B)_\psi [q \, q] 
+ \< \id^{A' \rightarrow \hat{B}} : \rho^{A'} \>.
\end{eqnarray*}
This is a channel version of the FQSW, obtained
by formally replacing $S$ with $A'$! We refer to this phenomenon
as \emph{source-channel duality}.
In the case where $A$ is null, the inequalities reduce to 
the two equivalent formulations of Schumacher compression;
in general, however, the two  are incomparable.
This observation sheds new light on the mysterious mother-father
duality \cite{DHW}, as the mother and father inequalities stem from  
FQSW and FF, respectively.
\begin{figure}
 \vspace{4mm}
  \includegraphics[width=8.5cm]{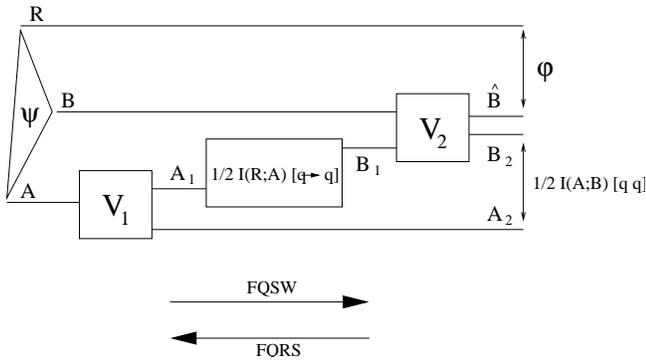}
  \caption{The protocol implementing FQSW consists of an isometry
$V_1: A \rightarrow A_1 A_2$ performed by Alice, sending $A_1$, 
of size $\lfloor n(1/2 \, I(R;A)_\psi + \delta) \rfloor$ qubits,
through a quantum channel $\id^{A_1 \rightarrow B_1}$ to Bob, 
and an isometry $V_2: B B_1 \rightarrow B_2 \hat{B}$ performed
by Bob; it approximately transforms $(\psi^{RAB})^{\otimes n}$ into 
$(\varphi^{R \hat{B}})^{\otimes n}$ and 
$\lfloor n/2 \, I(A;B)_\psi \rfloor  $ ebits shared between
Alice and Bob. The time reversal of this protocol implements the FQRS resource 
inequality.}
  \label{fig4}
\end{figure}

\paragraph{Duality \#3: FQRS is related to FQSW by time reversal.} 
We can make FQRS (\ref{RI1}) 
into a source type inequality by adding  
$\< \id^{S \rightarrow A'} : \rho \>$ to both sides of the equation:
\begin{eqnarray*}
  \< \id^{S \rightarrow A'} : \rho^S \>
+ 1/2  \, I(A;B)_\psi [q \, q] +  1/2  \,
I(R;B)_\psi [q \rightarrow q] \\
  \geq   \< \CU^{S \rightarrow AB} : \rho^S \>.
\end{eqnarray*}
Interchanging the roles of $A$ and $B$ gives
\begin{eqnarray*}
  \< \id^{S \rightarrow B'} : \rho \>
+ 1/2  \, I(A;B)_\psi [q \, q] + 1/2  \, I(R;A)_\psi [q \leftarrow q]  \\
  \geq   \< \CU^{S \rightarrow AB} : \rho^S\>.
\end{eqnarray*}
This is precisely the time-reversal of the FQSW inequality (\ref{RI3})! 
Unlike the previous two dualities, this one has \emph{operational} 
implications: 
a protocol achieving (\ref{RI1}) may be transformed into a protocol 
achieving (\ref{RI2}), and vice versa (figure \ref{fig4}).  

\begin{figure}
  \includegraphics[width=5cm]{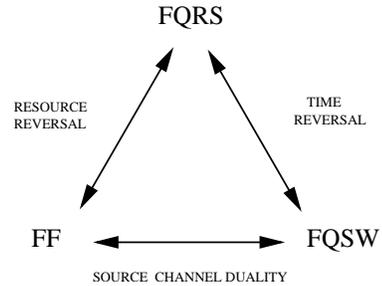}
  \caption{A triangle of dualities.}
  \label{fig2}
\end{figure}

Our three dualities are summarized in figure \ref{fig2}. 

\paragraph{The classical counterpart.} 
To what degree do these dualities carry over to classical information theory?
Let us define a classical (relative) feedback channel to take 
a random variable $X$ as Alice's input, and output a related random variable 
$Y$ to Bob, while feeding $XY$ back to Alice.
We shall use the simplified classical notation $\< X_A \rightarrow (XY)_A Y_B \>$ for such a 
resource. The same role played by purification in the quantum world is played by
copying in the classical world. 
Initially the reference system $R$ contains a copy of Alice's input state $X$;
whereas in the quantum case the feedback to $A$ was a purification of the $RB$ system,
here it is a classical copy of the $RB$ system. 
Notice the  breaking of ``purification symmetry'': 
while in the quantum case each of the $R, A$ and $B$ systems purifies the other two, 
here only $A$ is left with a copy of both $R$ and $B$.

The classical analogue of  FQRS is the classical reverse Shannon theorem \cite{BSST}
$$
I(X;Y) [c \rightarrow c] + H(Y|X) [c \, c] \geq 
\<  X_A \rightarrow (XY)_A Y_B \>.
\label{CRI1}
$$
The classical analogue of  FF is a feedback version of Shannon's channel coding theorem 
\cite{Sha}:
$$
\<  X_A \rightarrow (XY)_A Y_B\>  \geq 
I(X;Y) [c \rightarrow c] + H(Y|X) [c \, c].
\label{CRI2}
$$
We observe immediately that the resource reversal duality $\#1$ holds.

The classical analogue of FQSW is, not surprisingly, the original
Slepian-Wolf theorem \cite{SW}:
\begin{eqnarray*}
\<  (XY)_S \rightarrow X_A Y_B\> + H(X|Y) [c \rightarrow c]  \nonumber \\
  \geq  \<  (XY)_S \rightarrow (XY)_B \> .
\label{CRI3}
\end{eqnarray*}
The symmetry is now broken in a different way, with $R$ containing a copy
of $AB$, and there is no basis for dualities $\#2$ and $\#3$ to hold.

\paragraph{Conclusion.}
In summary, we have investigated three resource inequalities:
FQRS, FF and FQSW. These are implemented by variations on protocols 
exhibited elsewhere, via the adding of feedback or placing restrictions on 
channel inputs. All three involve only \emph{closed} quantum resources,
meaning that there is no mixing with an unobserved environment, but rather 
non-trivial distribution of quantum information among the protagonists. 
With this simplification, a beautiful structure emerges (figure \ref{fig2}):
FF and FQRS are related by the resource reversal
duality \#1; FF and FQSW are related by the source-channel duality \#2;  
FQRS and FQSW are related by the time reversal duality \#3.
Along the way we provide insights into: the GHZ decomposition problem \cite{LPSW}, 
the origin of the ubiquitous halves of quantum mutual informations, 
the difference between source and channel coding, the mother-father duality 
\cite{DHW}, and the breaking of ``purification symmetry'' in classical information theory.

\paragraph{Acknowledgments.}
We are grateful to T. Brun, P. Hayden, A. Harrow, A. Winter and J. Yard  
for useful comments on the manuscript.

\end{document}